\documentclass[a4paper,11pt]{article}
\usepackage{pos}

\usepackage{soul}
\usepackage{float}
\usepackage[
  backend=biber,
  style=numeric,
  sorting=none,
  giveninits=true,
  maxbibnames=99,
  doi=true,
  eprint=true,
  url=false,
  isbn=false
]{biblatex}

\DeclareFieldFormat[article,inproceedings,incollection]{title}{\mkbibquote{#1}}
\DeclareFieldFormat{journaltitle}{\mkbibemph{#1}}
\DeclareFieldFormat{booktitle}{\mkbibemph{#1}}

\newcommand{\alphas}{\ensuremath{\alpha_{\mathrm{s}}(M_{Z})}}

\newcommand{\MS}{\ensuremath{\overline{\mathrm{MS}}}}

\begin{document}

\title{Towards determination of the strong coupling $\alpha_s(m_Z)$
from four-flavor lattice QCD using the continuous $\beta$-function method}
\ShortTitle{Determination of $\alpha_s(m_Z)$ using the continuous $\beta$-function method}
 
\author*[a,b]{Yash Mandlecha}
\author[b,a]{Alexei Bazavov}
\author[c,d]{Akhil Chauhan}
\author[c,d]{Mingwei Dai}
\author[e]{Carleton DeTar}
\author[c,d]{Aida El-Khadra}
\author[f]{Steven Gottlieb}
\author[g]{Anna Hasenfratz}
\author[f]{Leon Hostetler}
\author[h]{Andreas S. Kronfeld}
\author[g]{Ethan T. Neil}
\author[b,a]{Curtis T. Peterson}
\author[h]{James Simone}
 
\affiliation[a]{Department of Physics and Astronomy, Michigan State University, East Lansing, MI 48824, United States}

\affiliation[b]{Department of Computational Mathematics, Science, and Engineering (CMSE),\\
Michigan State University, East Lansing, MI 48824, United States}

\affiliation[c]{Department of Physics, University of Illinois Urbana–Champaign, Urbana, IL 61801, United States}
\affiliation[d]{Illinois Center for Advanced Studies of the Universe, University of Illinois, Urbana, IL 61801, United States}

\affiliation[e]{Department of Physics and Astronomy, University of Utah, Salt Lake City, UT 84112, United States}

\affiliation[f]{Department of Physics, Indiana University, Bloomington, IN 47405, United States}

\affiliation[g]{Department of Physics, University of Colorado Boulder, Boulder, CO 80309, United States}

\affiliation[h]{Theory Division, Fermi National Accelerator Laboratory, Batavia, IL 60510, United States}

\emailAdd{mandlec1@msu.edu}

\abstract{
\vspace*{-2mm}
\textbf{\textsf{Fermilab Lattice and MILC Collaborations}}\\[0.7em]
The precise value of the strong coupling $\alpha_s(m_{Z})$ at the $Z$-boson mass $m_{Z}$ is essential for high-energy phenomenology and precision tests of quantum chromodynamics (QCD). We present the status of a program targeting a $\sim 0.3\%$ determination of $\alpha_s(m_{Z})$ using the renormalization group $\beta$-function in the infinite volume gradient flow scheme based on lattice QCD simulations of degenerate four-flavor highly improved staggered quark (HISQ) ensembles. In particular, we analyze both tree-level cutoff effects and finite-mass effects. We also outline the next steps of the analysis, including the infinite-volume and continuum extrapolations required for a precise determination of $\alpha_s(m_Z)$.
}

\FullConference{The 42nd International Symposium on Lattice Field Theory (LATTICE2025)\\
2-8 November 2025\\
Tata Institute of Fundamental Research, Mumbai, India\\}

%% \tableofcontents

\maketitle
\section{Introduction}

The strong coupling $\alpha_{s}\equiv \alpha_{s}(m_{Z})$, defined at the $Z$-boson pole mass ($m_{Z}$), is a fundamental parameter of the Standard Model. Its uncertainty feeds directly into precision determinations of the top-quark mass and into calculations of Higgs boson production rates at hadron colliders~\cite{Snowmass2021TF05, Snowmass2022WhitePaper}.   
It is also the dominant source of uncertainty in theoretical predictions of hadronic $Z$-boson decay widths~\cite{PDG2024}.
At present, lattice quantum chromodynamics (QCD) yields the most precise determination of $\alpha_{s}$, with a remarkable combined uncertainty of only about $0.6\%$ according to the 2024 report of the Flavor Lattice Averaging Group (FLAG 2024)~\cite{FlavourLatticeAveragingGroupFLAG:2024oxs}.
This already represents roughly a $33\%$ decrease 
in the uncertainty relative to the FLAG 2021 result~\cite{FlavourLatticeAveragingGroupFLAG:2021npn}. However, high-precision studies of Standard Model processes based on collider data require an uncertainty well below $1\%$, and the phenomenology community is currently aiming for a precision of $\lesssim 0.2\%$~\cite{Snowmass2021TF05, Snowmass2022WhitePaper}.

This project aims at a precise determination of $\alpha_{s}$ in the  infinite volume
gradient-flow scheme using the continuous $\beta$-function lattice approach~\cite{Luscher:2009eq, Luscher:2010iy, Fodor:2017die, Hasenfratz:2019hpg, Hasenfratz:2023bok, Wong:2023jvr}. Our computations employ a mass-degenerate four-flavor highly improved staggered quark action~\cite{Follana:2006rc}. Guided by the precision obtained with this method in the pure Yang-Mills  system~\cite{Hasenfratz:2023bok}, we aim for an uncertainty roughly a factor of two smaller than the FLAG 2024 global average. In these proceedings, we present the current status of our work.

We start in Sec.~\ref{sc} with an overview of the method and strategy used to determine the strong coupling. The ensembles employed in this work are summarized in Sec.~\ref{ensembles}. In Sec.~\ref{tln}, we discuss the impact of removing tree-level discretization effects on  the renormalized gradient flow coupling. Section~\ref{chiral} presents preliminary results for the chiral extrapolation at our largest bare couplings. Together, these steps form the initial core of our analysis pipeline. In Sec.~\ref{summary}, we summarize the progress achieved so far and describe the planned next stages of the study.

\section{Overview}\label{sec:overview}
\subsection{Strong coupling \boldmath$\alpha_s(m_Z)$ from gradient flow}
\label{sc}

The renormalization group (RG) $\beta$-function, $\beta(g^2)$, encodes how the renormalized coupling $g^2(\mu)$ varies with the energy scale $\mu$,
\begin{equation}\label{eqn:beta-function}
    \mu^2\frac{{\rm d} g^2(\mu)}{{\rm d}\mu^2} \equiv \beta(g^2).
\end{equation}
The $\Lambda$-parameter of QCD can then be obtained directly from the $\beta$-function by integrating Eq.~(\ref{eqn:beta-function}) from the ultraviolet fixed point at $g^2=0$ up to a reference scale $g^2 = g^2(\mu)$, yielding
\begin{equation}\label{eqn:lambda_parameter}
    \frac{\Lambda^2}{\mu^2} = \left(b_0 g^2(\mu)\right)^{b_1/b_0^2}\exp\left(-\frac{1}{b_0 g^2(\mu)}\right) \times
    \exp{\left[-\int^{g^2(\mu)}_0 {\rm d}x\left(\frac{1}{\beta(x)} + \frac{1}{b_0 x^2} - \frac{b_1}{b_0^2 x}\right)\right]},
\end{equation}
where $b_0$ and $b_1$ are the universal one- and two-loop coefficients of the RG $\beta$-function. A crucial physical ingredient in determining the $\Lambda$-parameter is the choice of hadronic reference scale. The conversion of this $\Lambda$-parameter to the $\MS$ scheme proceeds via an exact one-loop matching relation. Finally, using $\Lambda_{\MS}$ together with the 5-loop $\beta$-function in the $\MS$ scheme~\cite{Herzog:2017ohr}, one extracts $\alphas$ from Eq.~(\ref{eqn:lambda_parameter}), employing perturbative decoupling to translate the four-flavor $\beta$-function into its five-flavor counterpart.

We determine the $\beta$-function non-perturbatively using gradient flow, a continuous smearing transformation that evolves the fields along a fictitious flow time $\tau$
\cite{Luscher:2009eq, Luscher:2010iy}. The renormalized gradient-flow coupling is defined 
in a finite volume, with bare gauge coupling $\beta_{b} \equiv 10/g_{0}^2$, 
in terms of the flowed Yang–Mills energy density $E(\tau)$ as
\begin{equation}\label{eqn:coupling}
    g^2_{\rm GF}(\tau; L, \beta_{b})
    \equiv
    \mathcal{N}(a^2/\tau,L/a) \,
    \tau^2 \langle E(\tau) \rangle_{\beta_{b}}.
\end{equation}
The volume-dependent normalization factor $\mathcal{N}(a^2/\tau,L/a)$ serves several purposes: it ensures that the GF coupling matches the $\MS$ scheme at tree level in the combined continuum and infinite-volume limit; it accounts for gauge-field zero modes; and it can incorporate corrections to tree-level discretization effects~\cite{Fodor:2014cpa}. 
We investigate the inclusion of such tree-level improvement terms in more detail in Sec.~\ref{tln}. The expectation value $\langle \cdot \rangle_{\beta_{b}}$ denotes an ensemble average at fixed bare coupling $\beta_{b}$. The renormalization scale $\mu$ is related to the flow time by $\mu^{-1} \sim \sqrt{8\tau}$. The associated gradient-flow $\beta$-function is then defined as
\begin{equation}
    \beta_{\rm GF}(\tau;L,g_{0}^2)
    \equiv
    -\, \tau \, \frac{d g^2_{\rm GF}(\tau;L,g_{0}^2)}{d\tau},
\end{equation}
with the derivative evaluated using a five-point stencil. The infinite-volume and continuum limits are obtained using the continuous  (infinite volume) $\beta$-function method introduced in Refs.~\cite{Fodor:2017die, Hasenfratz:2019hpg, Hasenfratz:2023bok, Wong:2023jvr}. Briefly, this approach proceeds in two stages. First, at fixed $\beta_{b}$ and $\tau/a^2$, one takes the infinite-volume limit by extrapolating $g^2_{\mathrm{GF}}(\tau; L,\beta_{b})$ and $\beta_{\mathrm{GF}}(\tau; L,\beta_{b})$ linearly in $(a/L)^{4} \rightarrow 0$. Second, at fixed $g^2_{\mathrm{GF}}(\tau)$, the continuum limit is obtained by extrapolating $\beta_{\mathrm{GF}}(\tau; \beta_{b})$ to $a^2/\tau \rightarrow 0$. For ensembles in the confined or chirally broken phase, an additional extrapolation to the massless limit $am_{f} \rightarrow 0$ must be performed prior to taking the infinite-volume limit; we analyze this chiral extrapolation in Sec.~\ref{chiral}. In these proceedings, we consider only a single lattice volume and therefore defer both the infinite-volume and continuum extrapolations to future studies. 
\subsection{Ensembles}
\label{ensembles}
Our study employs gauge ensembles generated with four mass-degenerate fermions using the highly improved staggered quark (HISQ) action together with the tree-level Symanzik-improved (Lüscher–Weisz) gauge action~\cite{Luscher:1984xn, Luscher:1985zq}. We investigate twelve different values of $\beta_b$ spanning the interval $7.0 \leq \beta_b \leq 20.0$. 
For $8.0 \leq \beta_b \leq 20.0$,
the ensembles are massless ($am_f = 0$), whereas for 
%$7.00 \leq \beta_b \lesssim 7.50$ 
$7.0 \leq \beta_b \leq 7.5$ 
we simulate with at least three nonzero fermion masses in the range $(1.0 \leq am_f \leq 5.0)\times 10^{-3}$ and extrapolate to the massless limit. We are currently generating ensembles with spatial volumes $(L/a)^3$ and temporal extent $2L/a$, with $32 \leq L/a \leq 64$ at weak coupling ($8.00 \lesssim \beta_b \leq 20.0$) and $32 \leq L/a \leq 48$ at strong coupling ($7.00 \leq \beta_b \lesssim 7.50$). In this proceedings, we restrict our attention to the most complete subset of data, namely the ensembles with $L/a = 32$. 

\begin{figure}
    \centering
    \includegraphics[width=\linewidth]{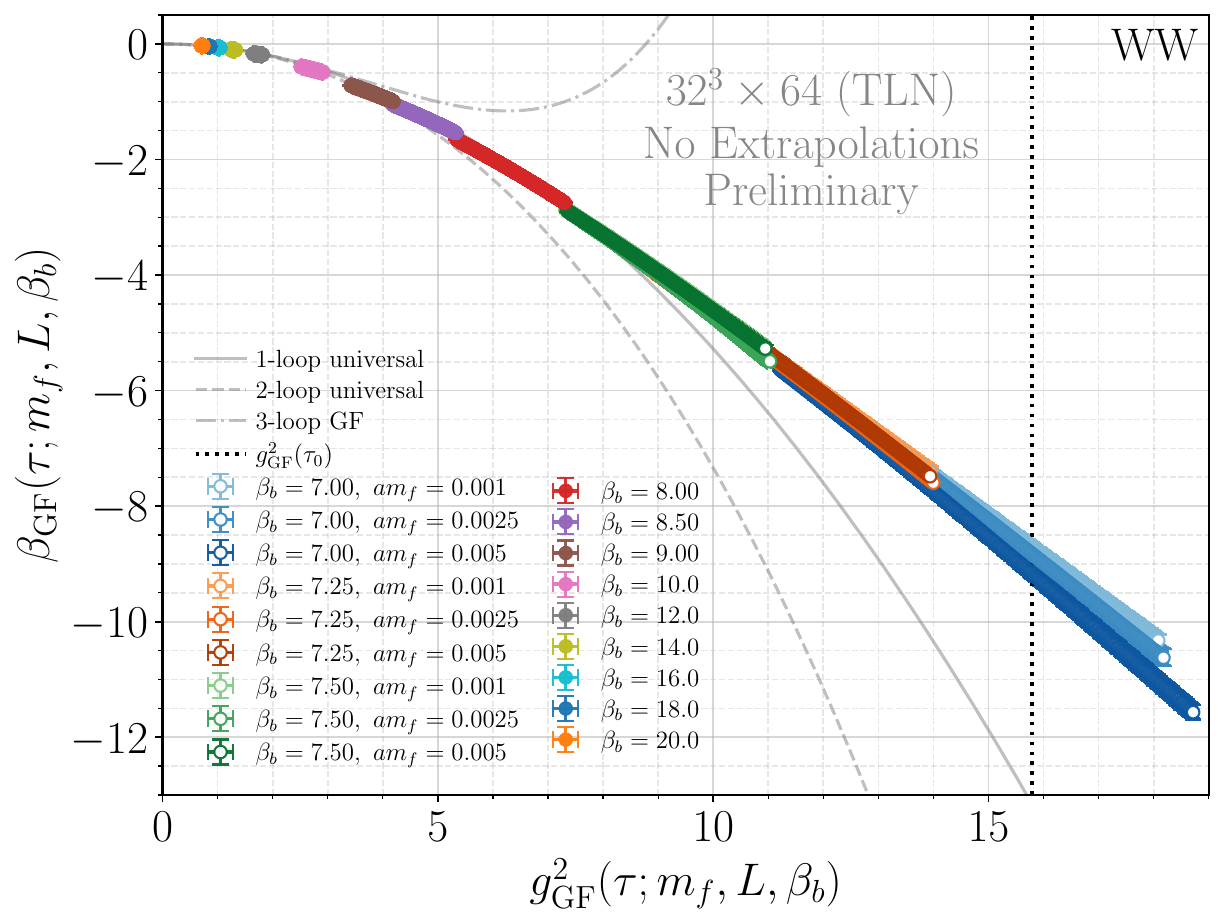}
    \caption{The gradient flow $\beta$-function $\beta_{\rm GF}(\tau; m_f, L, \beta_{b})$ on $32^3\times 64$ lattices is shown as a function of the running coupling $g^2_{\rm GF}(\tau; m_f, L, \beta_{b})$ over a broad range of bare couplings ($7.00 \leq \beta_{b} \leq 20.0$), spanning the weak- to strong-coupling regime $g^2_{\rm GF}\in(0.75,18)$.  
    Each color corresponds to a given bare gauge coupling $\beta_{b}$ and covers flow times $\tau/a^2 \in [2.0, 5.0]$. The width of each band represents the associated statistical error.  
    Simulations with $\beta_b\geq 8.0$ were carried out directly in the massless limit $am_f=0$, whereas the three strongest couplings ($\beta_{b}=7.00, 7.25$, and $7.50$) employ $am_{f} = 0.001, 0.0025$, and $0.005$, distinguished by different shades.
    All data have been corrected with tree level normlization (TLN). For comparison, we overlay the universal one-loop (solid) and two-loop (dashed) as well as the three-loop (dash-dotted) perturbative GF-scheme $\beta$-functions from Ref.~\cite{HarlanderNeumann2016}, all shown as gray curves.  
    The renormalized coupling at the reference scale $\tau_{0}$   ($g^2_{\rm GF}(\tau_0) = 0.3 \times 16 \pi^2/3 \approx 15.79$), is marked by the black vertical dashed line.}
    \label{beta_GF_raw}
\end{figure}
Gradient flow measurements are performed using Wilson (W) flow.
For the Yang-Mills energy density, three discretizations are considered: the Wilson (W), clover (C), and tree-level Symanzik (S) operators. We denote the flow ($X$) and discretization ($Y$) combinations as $XY$; for example, Wilson flow with Wilson discretization of the Yang-Mills energy density is WW.
Figure~\ref{beta_GF_raw} shows our prediction for the gradient flow $\beta$-function on $L/a=32$ over $7.00 \leq \beta_{b} \leq 20.0$ against the running gauge coupling, with each bare coupling indicated by a fixed color. Our result for the $\beta$-function is juxtaposed against the one- and two-loop universal $\beta$-function and perturbative three-loop gradient flow $\beta$-function~\cite{HarlanderNeumann2016}. The rightmost three curves corresponding to $\beta_{b}=7.00, 7.25$, and $7.50$ have $am_{f} = (1.0, 2.5, 5.0) \times 10^{-3}$, while all other bare gauge couplings have $am_{f}=0$. We observe that for the $L/a = 32$ ensembles, the $\beta$-function deviates from the perturbative one-loop prediction in the non-perturbative strong-coupling regime. We plan to calculate the $\Lambda$-parameter by integrating the $\beta$-function up to the $\tau_{0}$ scale, defined by $\tau_{0}^2\langle E(\tau_{0})\rangle \equiv 0.3$~\cite{Luscher:2010iy}. We indicate the location of $\tau_{0}$ in Figure~\ref{beta_GF_raw} by a vertical black dotted line. Our numerical results cross the $\tau_{0}$ scale, indicating that we've simulated at sufficiently strong couplings. To ensure that the continuum $\beta$-function sufficiently covers the region in the vicinity of $\tau_{0}$, however, we plan to add an additional bare gauge coupling in the $7.00 \leq \beta_{b} \leq 7.25$ region. 

\begin{figure}
    \centering
    \includegraphics[width=0.8\linewidth]{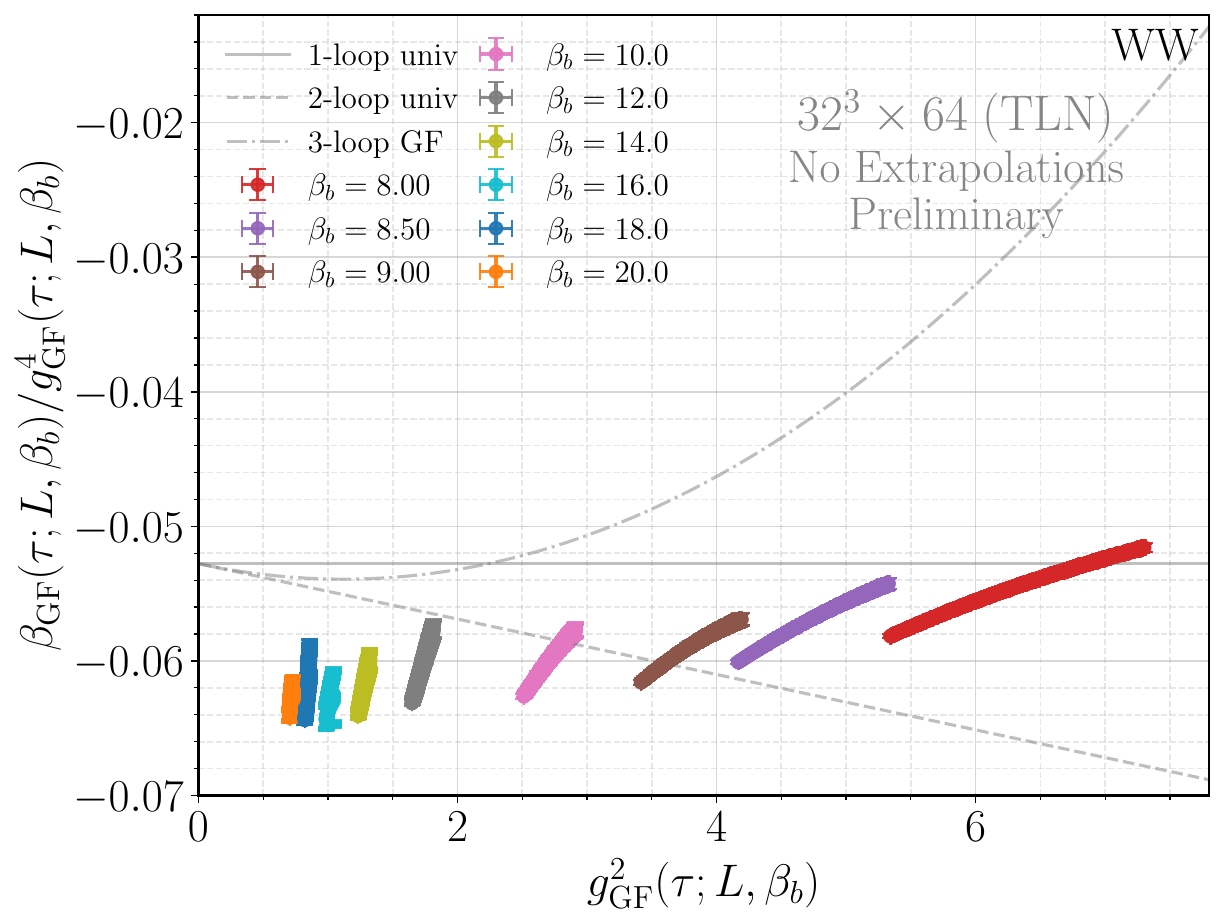}
    \caption{The gradient-flow $\beta$-function, $\beta_{\rm GF}(\tau; L, \beta_{b})$, divided by $g^4_{\rm GF}(\tau; L, \beta_{b})$, on $L/a=32$ volume as a function of the running coupling $g^2_{\rm GF}(\tau; L, \beta_{b})$ over the bare-coupling range $8.00 \leq \beta_{b} \leq 20.0$. Each color corresponds to a fixed bare gauge coupling $\beta_{b}$. For a given color (i.e., fixed $\beta_{b}$), the points trace out a curve as the flow time is varied over $\tau/a^2 \in [2.0, 5.0]$. The width of each band represents the associated statistical error. All data sets are obtained at $am_{f}=0.0$ and include tree-level normalization (TLN) corrections. The resulting $\beta$-function is compared with the universal one-loop (solid) and two-loop (dashed) perturbative predictions, as well as the three-loop (dash-dotted) perturbative $\beta$-function~\cite{HarlanderNeumann2016}, all shown as gray curves.}
    \label{betaGF_over_g4}
\end{figure}

As the renormalized coupling approaches the UV fixed point at $g^2_{\mathrm{GF}}=0$, our non-perturbative $\beta$-function should converge to the perturbative prediction. Figure~\ref{betaGF_over_g4} illustrates its behavior at our weakest couplings. To emphasize the weak coupling regime,   
we have divided the $\beta$-function by $g^4_{\mathrm{GF}}$.
The color scheme for each bare gauge coupling matches that of Figure~\ref{beta_GF_raw}. The data in this regime are especially significant, since they provide the dominant contribution to the $\beta$-function. 
The results shown in Figure~\ref{betaGF_over_g4}  appear to deviate from the perturbative curves even at weak coupling. This deviation is because we have not yet taken the $(a/L)^4 \to 0$ infinite volume and $a^2/\tau \to 0$ continuum limits. 
To control this crucial region of the $\beta$-function, we are currently generating data with lattice sizes up to $L/a=64$. 
\begin{figure}[H]
  \centering
  \begin{minipage}{0.5\textwidth}
    \centering
    \includegraphics[width=\linewidth]{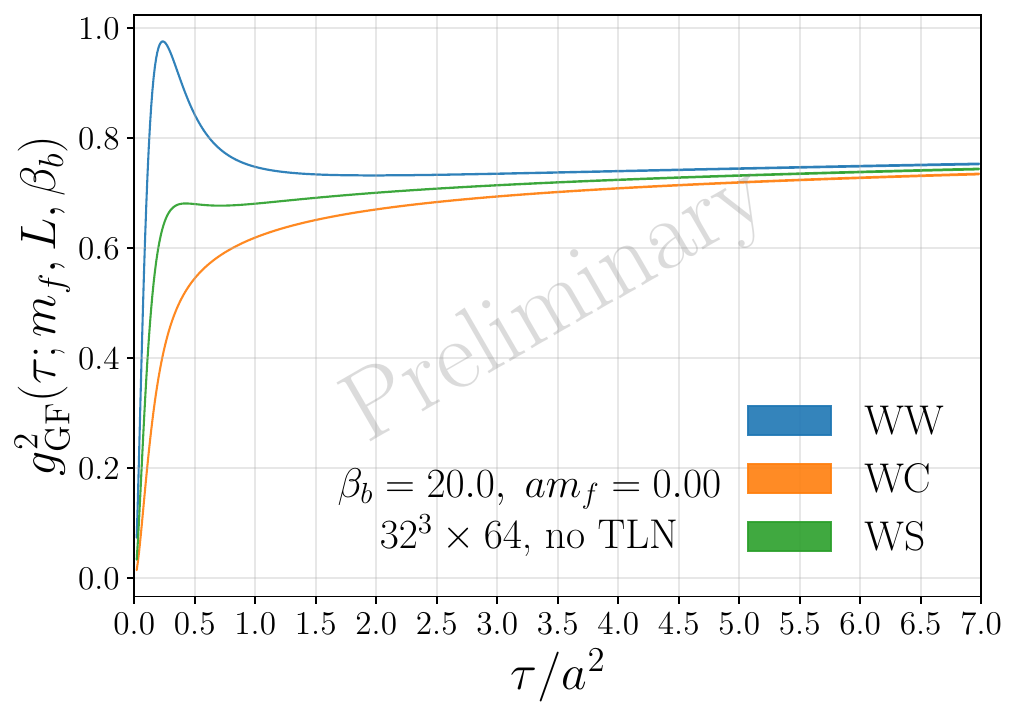}
  \end{minipage}\hfill
  \begin{minipage}{0.5\textwidth}
    \centering
    \includegraphics[width=\linewidth]{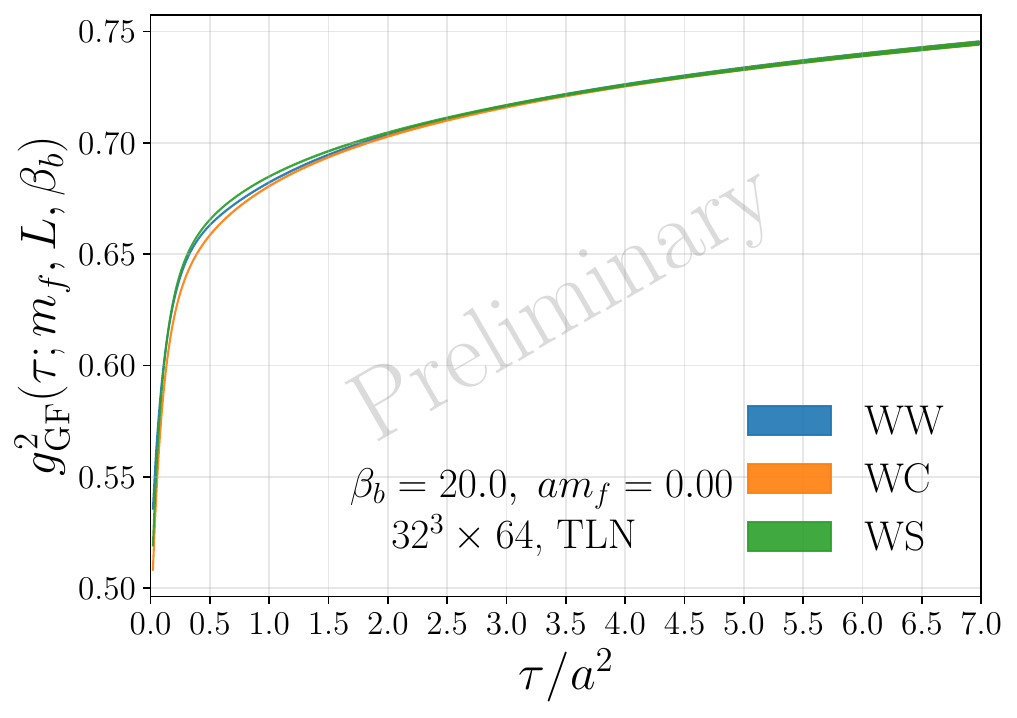}
  \end{minipage}
  \caption{\small
   Gradient flow running coupling $g^2_{\mathrm{GF}}(\tau;m_{f},L,\beta_{b})$ against the flow time in lattice units $\tau/a^2$ at $\beta_{b}=20.0$, $am_f=0.00$ without tree-level corrections (no TLN, left panel) and with tree-level corrections (TLN, right panel). Each color represents a fixed flow-discretization combination: WW (blue), WC (orange), and WS (green). 
  }
  \label{tln_weak}
\end{figure}
\begin{figure}[H]
  \centering
  \begin{minipage}{0.5\textwidth}
    \centering
    \includegraphics[width=\linewidth]{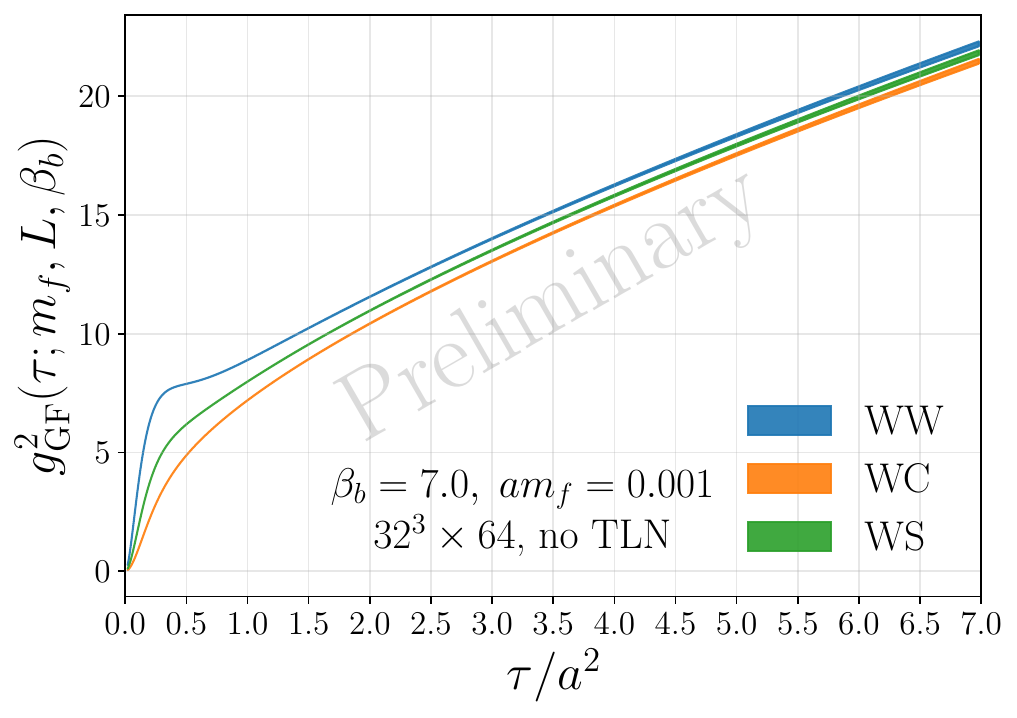}
  \end{minipage}\hfill
  \begin{minipage}{0.5\textwidth}
    \centering
    \includegraphics[width=\linewidth]{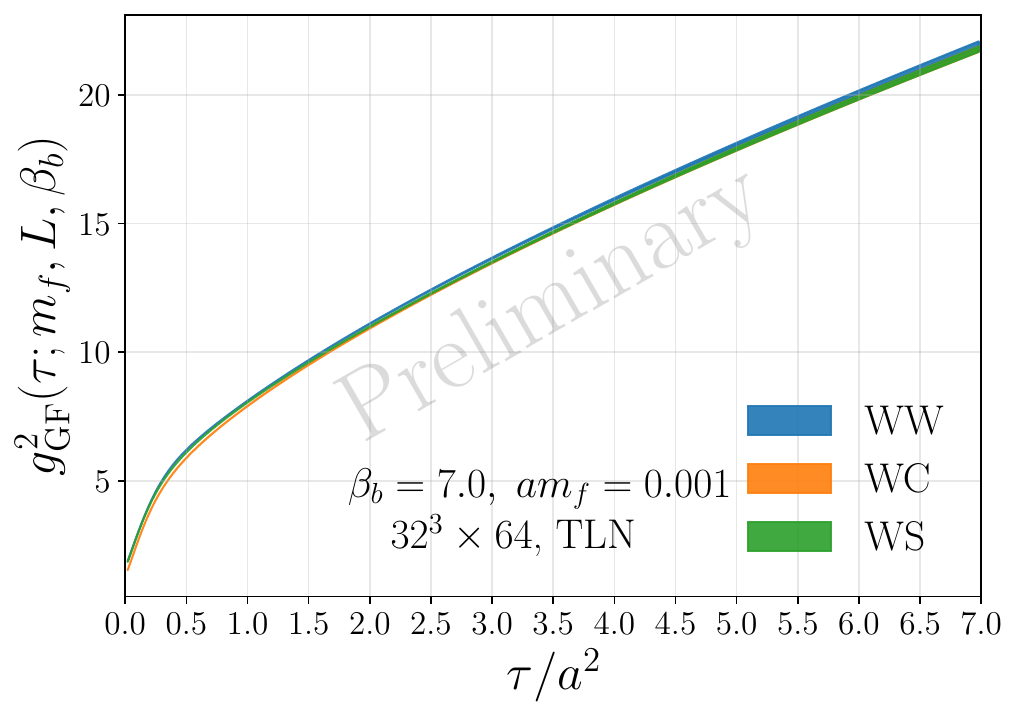}
  \end{minipage}
  \caption{
  Same as Figure~\ref{tln_weak} but at $\beta_b=7.0$, $am_f = 0.001$. The lattice data  are extrapolated to the massless limit.
  }
   \label{tln_strong}
\end{figure}
\section{Tree-level cutoff effects}
\label{tln}
At nonzero lattice spacing, discretization artifacts distort the expectation value of the gradient-flow energy density. Specifically, the lattice evaluation of the flowed observable differs from its continuum expression already at leading order in perturbation theory. This effect is described by
\begin{equation}
    \langle \tau^2 E(\tau) \rangle_{\rm latt} =
   \frac{3(N_c^2-1)}{128 \pi^2} g_0^2 \, 
   \left(C\Big(\frac{a^2}{\tau}, \frac{a}{L}\Big) + \mathcal{O}(g_0^2)\right),
\end{equation}
where the function $C(a^2/\tau, a/L)$ encapsulates lattice artifacts originating from both the finite lattice spacing $a$ and the finite spatial extent $L$. In the combined continuum and infinite-volume limits, $a^2/\tau \to 0$ and $a/L \to 0$, this correction factor tends to unity, reproducing the expected continuum tree-level behavior. By incorporating $C(a^2/\tau, a/L)$ into the normalization $\mathcal{N}(a^2/\tau,L/a)$ of Eq.~(\ref{eqn:coupling}), we eliminate all tree-level cutoff effects~\cite{Fodor:2014cpa}. We refer to the inclusion of $C(a^2/\tau, a/L)$ in $\mathcal{N}(a^2/\tau,L/a)$ as \textit{tree-level normalization}.
Figure~\ref{tln_weak} compares the gradient-flow coupling $g^2_{\rm GF}(\tau; m_f, L, \beta_b)$ as a function of $\tau/a^2$ for the weakest coupling $\beta_b = 20.0$ both with (right panel) and without (left panel) tree-level normalization. Results are shown for the Wilson, clover, and tree-level Symanzik discretization of the energy density. Differences in the renormalized coupling for different discretizations of the Yang-Mills energy density are indicative of cutoff effects. After application of the tree-level corrections, the gradient flow couplings from all three discretizations are more consistent, indicating a significant reduction in cutoff effects.
Figure~\ref{tln_strong} shows the same comparison for the strongest coupling considered, $\beta_b = 7.00$ with $am_f = 0.001$. Similar improvements were demonstrated in other systems~\cite{Hasenfratz:2023bok,Schneider:2022wmb}.
\section{Chiral Extrapolation}\label{chiral}
 \begin{figure}
    \centering
    \includegraphics[width=\linewidth]{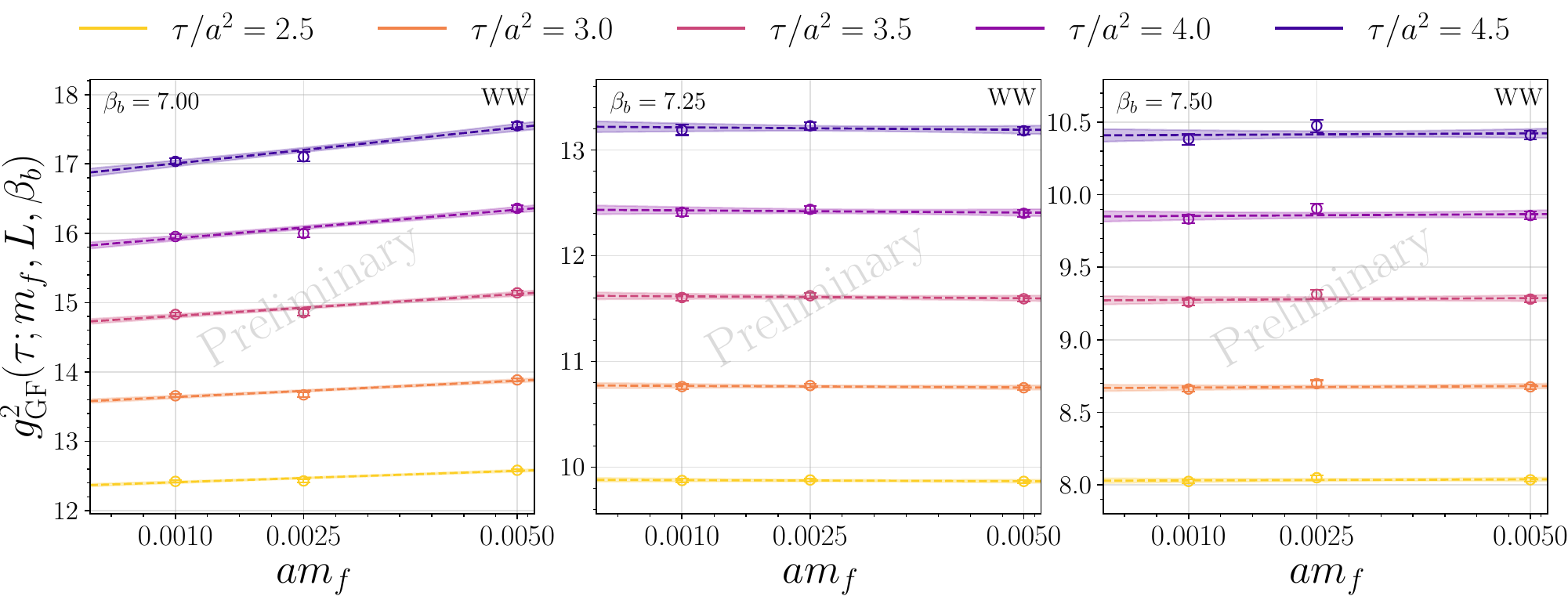}
    \caption{Chiral ($am_{f} \rightarrow 0$) extrapolation of the renormalized coupling $g^2_{\rm GF}(\tau; m_f, L, \beta_b)$ in $am_{f}$ at fixed flow times for strong bare couplings. Left panel shows $\beta_{b}=7.00$, middle panel $\beta_{b}=7.25$, and right panel $\beta_{b}=7.50$. Each color represents a fixed flow time $2.0\leq \tau/a^2 \leq 4.5$  (yellow to purple). The result of extrapolation indicated by a band, the width of the band indicating the statistical error and the central value of the band indicated by a dashed line. The data entering the extrapolation for each band is indicated by an error bar with a circular marker.}
    \label{g2_vs_mf_chiral}
\end{figure}
The RG $\beta$-function used to extract $\alpha_{s}$ is defined in the massless limit. Consequently, the renormalized coupling $g^2_{\mathrm{GF}}(\tau; L, \beta_b)$ must also be evaluated in the massless limit. 
In the weak-coupling, chirally symmetric, small-volume regime, simulations with massless quarks are feasible. In contrast, in the chirally broken, large-volume, strong-coupling regime, lattice simulations must be carried out at nonzero fermion mass and then extrapolated to the massless limit $am_f \to 0$. For sufficiently small fermion masses, the gradient-flow coupling is expected to depend linearly on $am_f$ at fixed flow time~\cite{Bar:2013ora, HarlanderNeumann2016}
\begin{equation}
   g^2_{\mathrm{GF}}(\tau; m_f, L, \beta_b)
   = g^2_{\mathrm{GF}}(\tau; 0, L, \beta_b)
   + A(\tau; L, \beta_b)\, am_f
   + \mathcal{O}(a^2m_f^2),
\end{equation}
where the slope $A(\tau; L,\beta_{b})$ depends on the bare coupling, flow time, and volume. We simulate our strongest couplings, namely $\beta_b = 7.00, 7.25$, and $7.50$, with $am_{f}=(1.0,2.5,5.0)\times 10^{-3}$.
Representative examples of the mass dependence and the corresponding chiral extrapolations at fixed flow time are shown in Figure~\ref{g2_vs_mf_chiral} for all three values of the bare coupling.
The left panel shows $\beta_{b}=7.00$, middle panel $\beta_{b}=7.25$, and right panel $\beta_{b}=7.50$. Each bare gauge coupling exhibits a mild dependence on the fermion mass, save for our strongest bare gauge coupling at $\beta_{b}=7.00$, for which the largest mass introduces a significant slope. It is worth noting that such behavior is not observed on our $L/a = 24$ dataset. 
\section{Summary and Future Prospects \label{summary}}
In this work, we presented progress toward a lattice determination of the RG $\beta$ function in the gradient-flow scheme, using ensembles that cover several lattice volumes and spacings. The renormalized running coupling was evaluated with multiple discretizations of the flowed energy density, which allowed us to investigate tree-level discretization effects. 
We also carried out a preliminary study of the chiral behavior of the renormalized coupling on our most strongly coupled ensembles. Together, these results represent essential initial steps toward a fully non-perturbative determination of the gradient-flow $\beta$-function.
The remaining stages of this project involve first carrying out an infinite-volume extrapolation to eliminate finite-size effects, and then performing a continuum extrapolation to obtain the fully renormalized running coupling and RG $\beta$ function in the continuum limit. To approach the infinite-volume limit, we are currently generating data on larger volumes for all bare gauge couplings. In parallel, we are performing a continuum-limit study of our finite-volume ensemble to better quantify the discretization effects that will remain once the infinite-volume limit is taken. After we have fully determined the $\beta$-function in the combined continuum and infinite-volume limits, we will extract the strong coupling following the procedure described in Sec.~\ref{sec:overview}. This determination of the strong coupling will be blinded in order to minimize bias. Upon completion of the project, we plan to release our complete gradient-flow dataset together with the analysis code and results underlying our main findings.
\section*{Acknowledgments}
We thank James Osborn and Xiao-Yong Jin for writing \texttt{Quantum EXpressions} (\texttt{QEX}) and helping us develop our \texttt{QEX}-based hybrid Monte Carlo and gradient flow software~\cite{Jin:2016ioq}.
 This work was supported by the Funding Opportunity Announcement Scientific Discovery through Advanced Computing: High Energy Physics, LAB 22-2580 (C.T.P., L.H.), the U. S. National Science Foundation under grant PHY23-10571 (C.D.).
Computations for this work were carried out in part with computing and long-term storage resources provided by the USQCD Collaboration. 
This work was supported in part by the U.S. Department of Energy, Office of Science, under Awards No.\ DE-SC0010005 (A.H., E.T.N.), No.\ DE-SC0010120 (S.G.), No.\ DE-SC0015655 (A.C., M.D., A.X.K.), the ``High Energy Physics Computing Traineeship for Lattice Gauge Theory'' No.\ DE-SC0024053 (A.C.), and the Funding Opportunity Announcement Scientific Discovery through Advanced Computing: High Energy Physics, LAB 22-2580 (L.H., C.T.P.); by the National Science Foundation under Grants Nos.\ PHY20-13064 and PHY23-10571 (C.D.), PHY23-09946 (A.B.), and Grant No.\ 2139536 for Characteristic Science Applications for the Leadership Class Computing Facility (L.H.).
A.H., A.X.K., A.S.K., and E.T.N. are grateful to the Kavli Institute for Theoretical Physics (KITP) for hospitality and support during the program ``What is Particle Theory?''
The KITP is supported in part by the National Science Foundation under Grant No.\ PHY-2309135.
A.X.K. and E.T.N. are grateful to the Pauli Center for Theoretical Studies and the ETH Zürich for support and hospitality.
This document was prepared by the Fermilab Lattice and MILC Collaborations using the resources of the Fermi National Accelerator Laboratory (Fermilab), a U.S.\ Department of Energy, Office of Science, HEP User Facility.
Fermilab is managed by Fermi Forward Discovery Group, LLC, acting under Contract No.~89243024CSC000002 with the U.S.\ Department of Energy. 

\printbibliography
\end{document}